\documentclass[%
 aip,
 amsmath,amssymb,
 reprint,%
]{revtex4-1}
\usepackage{graphicx, float}
\usepackage{dcolumn}
\usepackage{bm}

\usepackage[utf8]{inputenc}
\usepackage[T1]{fontenc}
\usepackage{mathptmx}
\usepackage{etoolbox}
\usepackage{lipsum}

\makeatletter
\def\@email#1#2{%
 \endgroup
 \patchcmd{\titleblock@produce}
  {\frontmatter@RRAPformat}
  {\frontmatter@RRAPformat{\produce@RRAP{*#1\href{mailto:#2}{#2}}}\frontmatter@RRAPformat}
  {}{}
}%
\makeatother
\begin{document}

\preprint{AIP/123-QED}

\title[The central dogma of biological homochirality: How does chiral information propagate in a prebiotic network?]{The central dogma of biological homochirality: How does chiral information propagate in a prebiotic network?}

\author{S. Furkan Ozturk}
\email{sukrufurkanozturk@g.harvard.edu}
\affiliation{Department of Physics, Harvard University, Cambridge, MA 02138, USA}

\author{Dimitar D. Sasselov}
\affiliation{Department of Astronomy, Harvard University, Cambridge, MA 02138, USA}

\author{John D. Sutherland}
\affiliation{MRC Laboratory of Molecular Biology, Cambridge, CB2 0QH, UK}

\date{\today}

\begin{abstract}
    Biological systems are homochiral, raising the question of how a racemic mixture of prebiotically synthesized biomolecules could attain a homochiral state at the network level. Based on our recent results, we aim to address a related question of how chiral information might have flowed in a prebiotic network. Utilizing the crystallization properties of the central RNA precursor known as ribose-aminooxazoline (RAO), we showed that its homochiral crystals can be obtained from its fully racemic solution on a magnetic mineral surface, due to the chiral-induced spin selectivity (CISS) effect \cite{ozturk2023origin}. Moreover, we uncovered a mechanism facilitated by the CISS effect through which chiral molecules, like RAO, can uniformly magnetize such surfaces in a variety of planetary environments in a persistent manner \cite{ozturk2023magnetization}. All this is very tantalizing, because recent experiments with tRNA analogs demonstrate high stereoselectivity in the attachment of L-amino acids to D-ribonucleotides, enabling the transfer of homochirality from RNA to peptides \cite{wu2021interstrand}. Therefore the biological homochirality problem may be reduced to ensuring that a single common RNA precursor (e.g. RAO) can be made homochiral. The emergence of homochirality at RAO then allows for the chiral information to propagate through RNA, then to peptides, and ultimately, through enantioselective catalysis, to metabolites. This directionality of the chiral information flow parallels that of the central dogma of molecular biology\textemdash the unidirectional transfer of genetic information from nucleic acids to proteins \cite{crick1958protein, crick1970central}.
\end{abstract}

\maketitle

\section{Introduction}

Biomolecular homochirality is a defining feature of life on Earth. While there is a solid understanding of its central role in biochemistry, the question of how homochirality could have possibly arisen is open, yet widely debated and explored ever since Pasteur's experiments on resolving chiral tartaric acid crystals \cite{pasteur1848relations}. As such it is a part of the bigger open question of life's origin, namely, of the prebiotic synthesis of the essential building blocks and the self-assembly of functioning polymers.

Common scenarios for how life might have emerged from prebiotic chemistry on Earth involve chemical synthetic networks. In chemical networks the flow of information can sometimes be quantified and used predictively \cite{wolf2012networks, schwaller2021networks}. However, the most successful applications to biology have historically been at a higher level of abstraction, especially in the face of fragmentary experimental results, as in Francis Crick's central dogma of molecular biology \cite{crick1958protein, crick1970central} (Fig. 1). Most work on the origin of biological homochirality has focused on discovering symmetry breaking chiral agents that act effectively on a single type or class of molecules. Less attention has been given to robustly propagating homochirality through an emerging biochemical network, partly because experiments on networks are difficult and inconclusive when enantiomeric excesses are small and amplification mechanisms are not well-matched and persistent \cite{blackmond2019origin, sallembien2022possible}. 

Recent findings on strong stereoselectivity in the attachment of amino acids to tRNA analogs \cite{wu2021interstrand}, combined with an RNA precursor capable of achieving robust and persistent homochirality by crystallizing on magnetized magnetite (Fe\textsubscript{3}O\textsubscript{4}) surfaces \cite{ozturk2023origin,ozturk2023magnetization}, provide for the first time an opportunity to demonstrate a prebiotically plausible pathway to network-level homochirality. In this paper we take a step further and use these experimental results to suggest that the homochirality achieved at the RNA precursor can propagate through RNA, then to peptides, and eventually to metabolites (Fig. 3). Thus the direction of chiral information flow in the developing prebiotic network parallels the directionality of genetic information propagation.

\section{Origin of Biomolecular Homochirality}

\begin{figure}
    \includegraphics[width=0.48\textwidth]{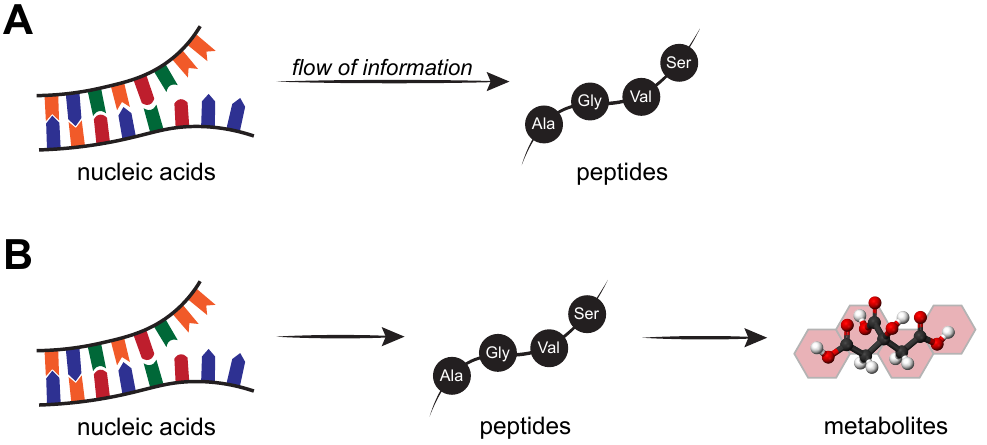}
    \caption{\textbf{A)} The central dogma of molecular biology, as established by Francis Crick, states that once the information has passed into proteins, it cannot get out. Therefore, the flow of genetic information is unidirectional, from nucleic acids to peptides. \textbf{B)} The central dogma can be expanded to include another molecular class: metabolites, because a metabolic network comprises compositional information determined by nucleic acids and peptides.}
    \label{Fig.1}
\end{figure}

Biological systems are homochiral, that is to say that those elements of their componentry which are chiral tend to be right-handed (D-) or left-handed (L-). Thus, for example, ribonucleic acid (RNA) is composed of ribonucleotides based on the sugar D-ribose and proteins are composed of L-amino acids\textemdash the two classes of molecules being said to have opposite relative stereochemistries. Conversely, individual metabolites can be of either handedness or achiral; the citric acid cycle, for example, involves L-malate, D-isocitrate, and a handful of achiral compounds. But how did this homochirality become established at the dawn of life?

Broadly speaking, and with respect to the prebiotic synthesis of the building blocks, two discrete mechanisms can be distinguished. In the first, each and every chiral building block or metabolite is individually resolved before incorporation into nascent biochemistry; in the second, one chiral building block or metabolite is resolved and then its homochirality ultimately controls the handedness of all others. Were there to be a generic way to resolve many chiral compounds in the same direction, such that all chiral amino acids were furnished left-handed, for example, then the first mechanism might be plausible, but this does not appear to be the case. The chemistry of nature’s score of proteinogenic amino acids is highly idiosyncratic and the chiral majority cannot be resolved in the same direction in a common way. The second mechanism is thus implicated, but this raises three major questions: What was the first chiral protobiological compound to be resolved?, How did it get resolved?, and What trajectory was followed to allow its handedness to control the relative stereochemistry of other chiral compounds and thus allow homochirality to propagate throughout the developing biological network? 

The primary purpose of this article is to answer the third question, but as it is ineluctably connected to the other two, we must also address them. Before we do so, it is instructive to consider chirality in the context of molecular information. 

\section{Central Dogma of Biological Homochirality: Direction of Chiral Information Flow}

The idea of chirality propagation during the development of a biological system is then highly reminiscent of Crick’s concept of unidirectional information transfer at the heart of biology \cite{crick1970central}. His central dogma has it that information written in the language of nucleic acid sequence is irreversibly translated into information written in the different language of protein sequence \cite{crick1958protein} (Fig. 1A). Recognizing that a metabolic network comprises compositional information and that the composition is controlled by nucleic acids and proteins, the central dogma can be extended to say that information in nucleic acids both directly, and indirectly through proteins, is converted into compositional metabolic information (Fig. 1B). Although subsequent nucleic acid and protein synthesis is provisioned by metabolites, compositional metabolic information is not passed on\textemdash the information in nucleic acids is propagated by replication. 

\begin{figure}
    \includegraphics[width=0.49\textwidth]{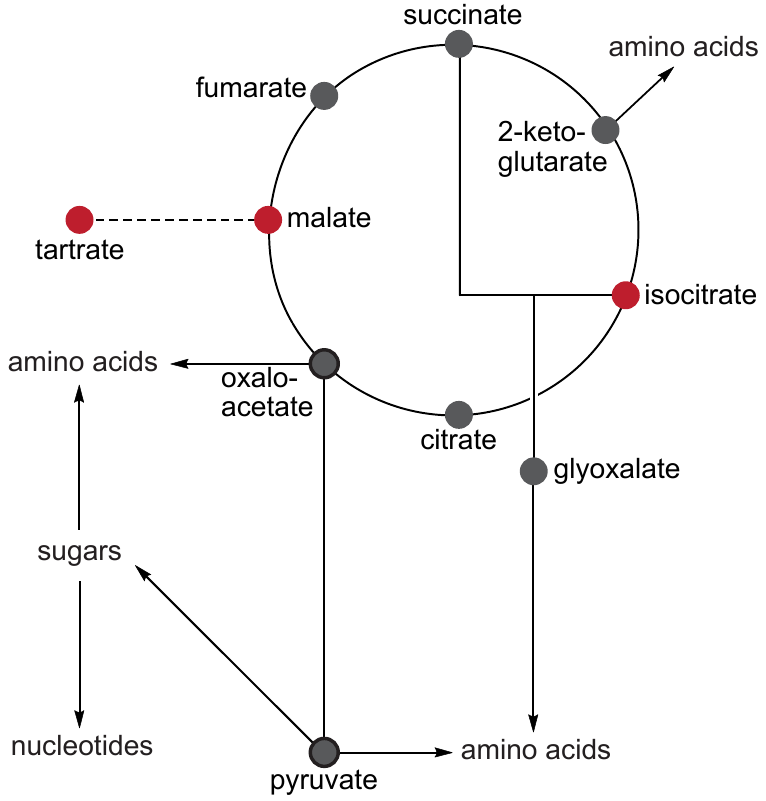}
    \caption{Core metabolism\textemdash absent catalysis by enzymes, cannot propagate chiral information. The two chiral entities (L-malate and D-isocitrate, shown in red) are flanked by achiral intermediates (shown in gray). Making matters worse, the achiral branch points (oxaloacetate and pyruvate) inhibit the flow of chiral information from/through the metabolites. The theoretical provisioning of metabolism using chiral tartrate as a precursor is indicated by the dashed line.}
    \label{Fig.2}
\end{figure}

What about the propagation of homochirality? Once homochirality is established for a certain compound, how could the chiral information efficiently spread into additional compounds in the same class and other classes of molecules? To answer this, we do not have to make assumptions about the origin of biomolecular homochirality; we can address the origin later, based on our latest results. For now, let's take as our starting point the homochiral state of a single chiral compound and inquire which molecular class (in Fig. 1B) would be the ideal one to establish and propagate homochirality to the entire prebiotic network in efficient and prebiotically plausible manner absent homochiral macromolecular catalysts.

\begin{figure*}
    \includegraphics[width=0.8\textwidth]{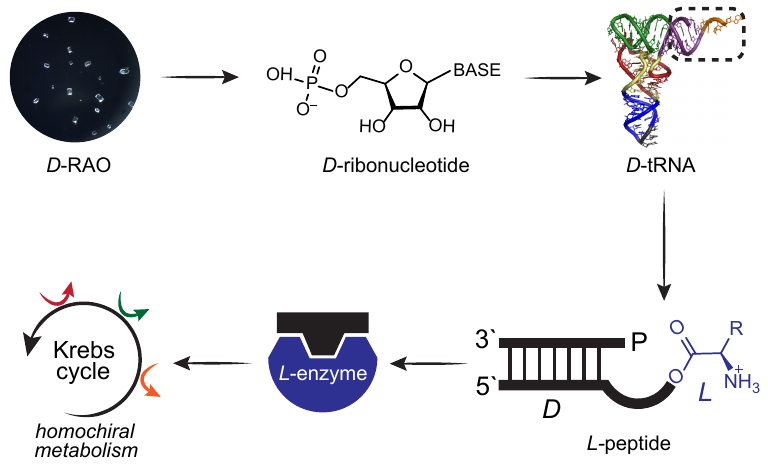}
    \caption{Propagation of homochirality from a common ribonucleotide precursor, RAO, to the entire prebiotic network is illustrated. D-RAO can be transformed into D-ribonucleotides, forming homochiral RNA. L-peptides can be synthesized through the stereoselective binding of L-amino acids to tRNA analogs composed of D-ribonucleotides. This crucial link enables the transfer of homochirality from nucleic acids to peptides. Then L-peptides form L-enzymes, and by enantioselective catalysis a homochiral metabolism can be produced.}
    \label{Fig.3}
\end{figure*}

\subsection{Chiral Information Flow from Metabolites}

Emergence of homochirality in metabolites is in principle possible. As we will detail the mechanism in the next section, crystallizing racemic tartrate salts on a magnetic surface can produce enantiopure L-tartrate, which can in principle be reduced to L-malate \cite{liu2021prebiotic}. However, when L-malate enters the citric acid cycle (Krebs cycle), it is either dehydrated to fumarate or oxidized to oxaloacetate, both of which are achiral compounds (Fig. 2, red dots). Homochirality established at D-isocitrate suffers the same fate. Therefore, due to the presence of achiral intermediates and branching points, the propagation of homochirality from metabolites\textemdash to their entire class or to amino acids and nucleotides\textemdash seems to be impossible (Fig. 2).

\subsection{Chiral Information Flow from Amino Acids}

Metabolic reactions in biology are carried out by enzymes which are composed of L-amino acids. Therefore, if amino acids are made homochiral, homochiral enzymes can synthesize the chiral metabolites enantioselectively and the chiral information can propagate from amino acids to metabolites. The achiral intermediates which constituted a barrier to propagation of chirality in the absence of enzymes can now be overcome. Here the chiral information flow follows Crick's expanded central dogma (Fig. 1B). 

However, the emergence of homochirality within the peptides themselves encounters a problem, because there are 19 chiral proteinogenic amino acids and they all have idiosyncratic chemistries. Although it is possible to resolve two conglomerate-forming amino acids (Asn, Glu.HCl) with moderate enantiomeric excess by crystallization on magnetic surfaces \cite{tassinari2019enantioseparation}, resolving all amino acids with different reaction pathways in the same direction is a tedious task, and has never been demonstrated experimentally in a prebiotically plausible manner. Moreover, propagating the chiral information inside the class remains an open question as well. While we do not rule out the possibility of an induced enantiomeric excess at a few amino acids being transferred to the rest, there is currently no experimental evidence to support this.

Finally, what about propagating chiral information from amino acids to nucleic acids? Experiments with proline and proline-valine peptides showed an ability to transfer enantiomeric excess to glyceraldehyde, a 3-carbon sugar and an RNA precursor \cite{hein2011route, yu2021prebiotic}. While the results with proline are highly promising, they remain an isolated case and the emergence of homochiral proline remains an open problem. If enantiopure proline can be obtained under prebiotic conditions through a process controlled merely by the environment, these results can have significant implications for the chiral information flow from amino acids to nucleic acids.

Though worth mentioning from the chemistry perspective, the results via the formose reaction and chiral catalysts are less relevant prebiotically \cite{breslow2010amino, burroughs2012asymmetric}. As a consequence, emergence of homochirality in amino acids and its subsequent propagation to other molecular classes\textemdash directly or through homochiral enzymes is probable though not fully supported by current experimental knowledge.

\subsection{Chiral Information Flow from Nucleic Acids}

What if peptides are synthesized from racemic amino acids by the genetic code, which is homochiral? 

Recent findings demonstrate that L-amino acids can be selectively attached to tRNA analogs composed of D-ribonucleotides with up to 10-fold stereoselectivity (Fig. 3) \cite{wu2021interstrand}. This chemistry involves attachment of an amino acid to a phosphate group at one terminus of an RNA chain and subsequent migration to the diol group at the other end of the chain via a looped conformation. The stereoselectivity is in the migration reaction: L-aminoacyl-residues are transferred faster than D-configured residues. Because the initial tether to the phosphate is hydrolytically-labile, the D-configured residues which are the slowest to transfer are predominantly removed by hydrolysis. Experimentally, this has been demonstrated using D- and L-alanyl-residues, but modelling suggests the stereoselectivity will be the same for other aminoacyl-residues \cite{wu2021interstrand}. 

Hence, a homochiral genetic molecule can facilitate the synthesis of peptides predominantly consisting of L-amino acids from a racemic pool of amino acids. As homochiral peptides can carry the chiral information to metabolites through enantioselective catalysis, homochirality established at the genetic molecule can effectively propagate into the entire prebiotic network\textemdash suggesting that the direction of chiral information flow parallels the directionality of information propagation (Fig. 1 vs Fig. 3). 

The implication here is that the overall problem of homochirality's origins may be reduced to the problem of making enantiopure ribonucleotides (Fig. 3). Consequently, if a common ribonucleotide precursor is obtained in its enantiopure form, an entire prebiotic network can then be made homochiral. One such precursor is ribose-aminooxazoline (RAO).

\section{Prebiotic Synthesis of Ribonucleotides and Ribose-aminooxazoline (RAO)}

That ribose-aminooxazoline might be an intermediate in prebiotic ribonucleotide synthesis was first recognized by Orgel \cite{sanchez1970studies}. Although he presciently commented on its tendency to crystallize, he did not find a high-yielding pathway from it to ribonucleotides. He also found no other way to make it other than by starting from ribose, a sugar that has only ever been made as a trace component of a complex mixture under prebiotically plausible conditions \cite{shapiro1988prebiotic}. Subsequently the Sutherland group discovered that RAO could be made starting from glycoaldehyde and glyceraldehyde and that these two sugars could in turn be easily made from hydrogen cyanide by reductive homologation in what was termed cyanosulfidic chemistry \cite{anastasi2006direct, ritson2013synthesis}. The same group also discovered that an anhydronucleoside derivative of RAO could be ring-opened by hydrogen sulfide and the resultant product photoepimerized and hydrolyzed to give the canonical pyrimidine ribonucleosides, cytidine and uridine \cite{xu2017prebiotically}. Routes from RAO to the purine ribonucleosides continue to be explored, but thus far an efficient synthesis of these nucleosides from RAO has not been described. 

RAO belongs to a small subset of crystalline chiral organic compounds. Whilst the majority of such compounds crystallize as racemic compounds with equal amounts of both enantiomers in all crystals, RAO crystallizes as a conglomerate in which individual crystals contain one or other enantiomer \cite{springsteen2004selective}. It transpires that this conglomerate behaviour is crucial to resolution of RAO \cite{ozturk2023origin}. It is because such behaviour is rare that efforts to find chemistry diverging from RAO to all the canonical ribonucleosides are continuing. 

\section{Chiral-Induced Spin Selectivity and Chemistry Controlled by the Electron Spin}

The chiral-induced spin selectivity (CISS) effect is a new phenomenon relating the electron spin to molecular chirality \cite{naaman2012chiral, naaman2019chiral}. Early observations of the CISS effect have revealed that electron transmission probability through a chiral system is highly spin dependent and the molecular chirality of the system directly determines the spin state of electrons that are favorably transferred \cite{ray1999asymmetric, gohler2011spin, naaman2019chiral}. As such, the CISS effect has established a robust coupling of the electron spin to the molecular chirality of chiral molecules. Due to this strong coupling, near unity spin polarizations have been observed in electron transmission experiments from a chiral self-assembled monolayer \cite{gohler2011spin, kettner2015spin}. Although a full theoretical description of the CISS effect is still an active search \cite{evers2022theory, fransson2019chirality}, qualitative origins of the observed spin polarization can be explained by an effective spin-orbit interaction \cite{varela2019intrinsic, naaman2012chiral, evers2022theory}. This effective spin-orbit interaction is due to a magnetic field an electron experiences in its own reference frame as it moves inside the chiral electrostatic potential of a chiral molecule.

While chiral molecules can be used to filter electrons depending on their spin state, the coupling established by the CISS effect can also be used to initiate chirally selective asymmetric processes that are controlled by the electron spin \cite{naaman2020chiral}. Because electron spin strongly couples to molecular chirality in an enantioselective manner, a process initiated or driven by electrons with a well-defined spin alignment can be made enantioselective. In other words, a bias in the (net) spin alignment can be used to induce enantiomeric excess, due to the CISS effect. Therefore, electron spin can function as a chiral agent and conduct chirally selective chemical processes \cite{bloom2020asymmetric, metzger2020electron, metzger2021dynamic}.

A natural question then is where to find a bias in the electron spin in a prebiotic setting. To address this question we recently proposed closed-basin evaporative lakes with sedimentary magnetite (Fe\textsubscript{3}O\textsubscript{4}) deposits as plausible prebiotic environments that can accommodate spin-selective processes \cite{ozturk2022homochirality}. In these environments, authigenic magnetite forms as single-domain, superparamagnetic particles and gets magnetized under the geomagnetic field \cite{harrison2002direct, hurowitzetal2017, toscaetal2018}. Later, these magnetite crystals settle on the lake bottom and carry a statistically uniform remanent magnetization across a hemisphere of the planet \cite{karlin1987authigenic}. The net magnetization (spin-polarization) of the sediments breaks the chiral symmetry and can trigger spin-selective processes near the magnetic surface, due to the CISS effect. Therefore, magnetic mineral surfaces magnetized by the planetary magnetic field can function as prebiotically available chiral agents that introduce a chiral bias to prebiotic chemistry through processes controlled solely by the physical environment.

\subsection{Magnetic Surfaces as Chiral Agents}

\begin{figure*}
    \includegraphics[width=0.85\textwidth]{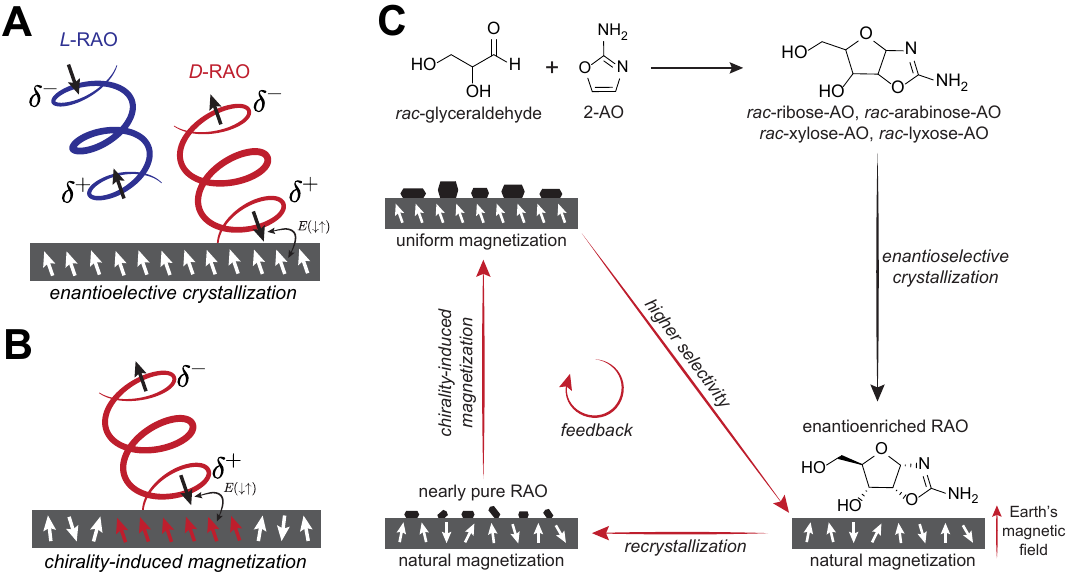}
    \caption{\textbf{A)} Due to the CISS effect, magnetic surfaces with net magnetization can function as chiral agents and be templates for the enantioselective crystallization of RAO from its racemic solution, by a spin-controlled process. \textbf{B)} Selective interaction of chiral molecules with the electron spin can be utilized to magnetize magnetic surfaces. Magnetic surfaces with no net magnetization can be uniformly magnetized by crystallization of enantiopure RAO, along the chiral molecular axis. \textbf{C)} Enantioselective crystallization of chiral molecules on magnetic surfaces can be considered together with the chirality-induced magnetization in a cooperative feedback. With this feedback, a small bias in the natural magnetization of magnetic surfaces due to the geomagnetic field can be amplified, and enantioselective processes controlled by the electron spin can be made highly efficient. Note that racemic RAO is synthesized with three other racemic aminooxazolines (arabinose-AO, xylose-AO, lyxose-AO), which get excluded in the initial crystallization step due to the relatively low solubility of RAO.}
    \label{Fig.4}
\end{figure*}

In our original proposal, we suggested using UV-ejected solvated electrons from the magnetized surface to induce reduction reactions near the surface where the electrons will have a chiral character due to the net alignment of their spin and momenta. However, these freely diffusing electrons are likely to lose their chiral character shortly after they leave the surface as their net alignment of momenta gets scrambled \cite{ozturk2022homochirality}. Further considerations of the low photoelectron yield of magnetite by UV irradiation shifted our attention from chiral electrons to magnetic surfaces themselves as agents that can impose a chiral bias on prebiotic chemistry.

Achiral magnetic surfaces with net magnetization can function as chiral agents due to the CISS effect and magnetic minerals can facilitate enantioselective surface processes (e.g. adsorption, crystallization) that are controlled by the net magnetization of the surface \cite{tassinari2019enantioseparation, banerjee2018separation, bhowmick2021simultaneous}. 

In its essence, the CISS effect is about the spin-selectivity of electron flow in a chiral medium such as the chiral electrostatic potential of a chiral molecule. This implies that by subjecting a chiral molecule to an electric field along its molecular axis, it is possible to achieve a spin-selective electron flow through the chiral molecule \cite{naaman2019chiral}. As a molecule approaches a surface, its electron density is reorganized and the molecule acquires an induced transient charge polarization due to the dispersive forces between the molecule and the surface. And because of the CISS effect, this induced transient charge polarization is accompanied by a transient spin-polarization such that one electric pole is associated with one spin state and the other pole with the opposite one \cite{naaman2019chiral, ziv2019afm}. The spin direction is defined along the chiral molecular axis\cite{meirzada2021long}, with the molecular handedness determining which electric pole is linked to each spin state (Fig. 4A). It is important to appreciate that while molecules tumble in a solution, when they come into proximity with a surface, they align themselves in relation to the surface. The specific arrangement of molecules and their angular orientation near the surface are determined by the chemical properties of the molecular moieties and the surface itself.

Now, suppose the surface has a net spin polarization (i.e. the surface is magnetized), what would happen? In this case, the surface spins interact with the transient unpaired spins of the chiral molecules through spin exchange interaction. The enantiomer forming a singlet-like lower energy spin configuration with the surface will be attracted to the surface more and the other enantiomer will be repelled as it forms a triplet-like spin configuration with a higher energy. As a consequence, the concentration of a particular enantiomer near the surface will be higher based on the surface's magnetization direction, leading to faster crystallization of that enantiomer analogous to chiral resolution by kinetic entrainment \cite{jacques1981enantiomers}. Although, the spin exchange interaction is short-ranged (< nm), it is a strong one. The energy difference between the singlet and triplet-like configurations ($\sim0.01-1$ eV) can be much higher than the room temperature and therefore the consequences of  spin exchange interaction can be observed in a robust way \cite{ziv2019afm, safari2022enantiospecific}. In summary, magnetized surfaces can function as templates for the asymmetric crystallization of chiral molecules and seed their crystallization in a chirally selective manner, determined by their magnetization direction.

\subsubsection{Spin-selective Crystallization of RAO on a Magnetic Surface}

We utilized magnetized magnetite (Fe\textsubscript{3}O\textsubscript{4}) surfaces as chiral agents that were potentially available in prebiotic settings and by crystallizing RAO from its fully racemic solution on these surfaces, we successfully obtained homochiral crystals of RAO in two crystallization steps \cite{ozturk2023origin} (Fig. 4A).

First, we synthesized racemic RAO together with other racemic pentose aminooxazolines (arabinose-AO, xylose-AO, and lyxose-AO) by the reaction of racemic glyceraldehyde and 2-aminooxazole (Figure 4C). Then, by a direct crystallization of this mixture, we exclusively obtained RAO crystals due to the low solubility of RAO compared to the other pentose aminooxazolines. This initial crystallization was done on a magnetic surface, however, only a minimal enantiomeric excess could be obtained when RAO was crystallized in the presence of xylose-AO (XAO)\textemdash a crystal habit modifier for RAO. However, when racemic RAO is re-crystallized on magnetized magnetite, from its own aqueous solution, without XAO, we obtained nearly enantiopure crystals of RAO (Fig. 4C). 

These experiments were done with an external magnet present, which was positioned in a way that the Fe\textsubscript{3}O\textsubscript{4} surface was strongly magnetized in an out-of-plane direction. Enantioselective crystallization of RAO on the magnetic surface was controlled by the magnetization direction of surface and by flipping the polarity of the external magnet, we could switch the handedness of the RAO crystals. However, it should be emphasized that the effect is due to a spin-exchange interaction between the spin-polarized surface and chiral RAO molecules, not a magnetic interaction. To verify this, we crystallized RAO on non-magnetic silicon surfaces in the presence of an external magnet and did not observe any enantioselectivity. 

As mentioned before, the spin-exchange interaction is a short-ranged one therefore the  surface spins could only influence the crystallization nearby the surface. However, because RAO crystallizes as a conglomerate, a surface-level enrichment could be propagated by the conglomerate crystallization of RAO. 

Nevertheless, it is essential to note that the magnetic surfaces utilized in the experiments had a near-uniform spin polarization, which was achieved through the application of a strong external magnetic field. Such a field would not be present in a prebiotic environment yet a near-uniform spin polarization can still be achieved by different means under prebiotic conditions. In a follow-up work, we addressed this issue and discovered a mechanism that enables magnetic surfaces to attain a nearly uniform magnetization at the surface via the use of chiral molecules \cite{ozturk2023magnetization}.

\subsubsection{Chirality-Induced Magnetization of Magnetic Surfaces}

In a natural environment, authigenic magnetite sediments form and magnetize under Earth's geomagnetic field. Although this authigenic magnetization is significant and statistically uniform on a planetary hemisphere scale, it is by no means uniform. Nonuniformity of the surface magnetization translates into a lower attained enantioselectivity in surface processes controlled by the electron spin. We have shown that chiral molecules can magnetize magnetic surfaces to a state of near uniformity due to the CISS effect \cite{ozturk2023magnetization}, and hence suggest that a large-scale and uniform magnetization can be attained under prebiotically plausible conditions.

The interaction responsible for the enantioselective magnetization of magnetic surfaces can also be employed in reverse, whereby chiral molecules align the spins of a magnetic surface through spin-exchange interaction. As shown in Fig. 4B, as a chiral molecule approaches a surface with no net magnetization, it flips the surface spins along its chiral molecular axis due to a singlet-like coupling between the transient chiral molecular spin and the surface spins. Therefore, as an enantiopure layer of a chiral molecule gets adsorbed on a surface with no net prior magnetization, the surface spins acquire a net magnetization whose direction is determined by the molecular handedness of the layer. We studied this process by crystallizing enantiopure RAO on magnetite surfaces with no net magnetization and measured the induced magnetization of magnetite by the chiral molecules. We observed that magnetic domains underneath and nearby the chiral crystals get magnetized along a common direction, dictated by the molecular chirality of the crystals \cite{ozturk2023magnetization}.

Moreover, the magnetization induced by the chiral molecules propagates like an avalanche, leading to a uniform magnetization that extends beyond the region covered by the RAO crystals. Our measurements of the surface magnetization also revealed that the surface magnetized by the chiral crystals has a higher magnetic resistance (coercivity) compared to a portion of the surface away from the chiral molecules \cite{ozturk2023magnetization}. The increase in the magnetic coercivity was measured to be about 1 mT which is about 20 times higher than the modern geomagnetic field, indicating that the induced surface magnetization will persist against possible geomagnetic reversals \cite{o2014evolution}. This finding has important implications for prebiotic chemistry, as it suggests that the same chiral bias would be maintained in the environment in a deterministic fashion, at different stages of prebiotic chemistry. 

In summary, just as magnetic surfaces influence chiral molecules, chiral molecules couple with the magnetic domains underneath them due to the spin-exchange interaction. This mutual coupling paves the way for a cooperative feedback effect.

\subsubsection{Feedback Between Chiral Molecules and Magnetic Surfaces}

Considered together with our previous results on enantioselective crystallization, the chirality-induced magnetization phenomenon verifies a reciprocal relationship between chiral molecules and magnetic surfaces: magnetic surfaces with net magnetization can induce enantioselective processes and attained enantiomeric excess can lead to an increase in the net magnetization. 

In our initial work on enantioselective crystallization, we found that a magnetite surface magnetized by the north (south) pole of a magnet promotes the crystallization of D-RAO (L-RAO). To our delight, in our follow-up work, we observed that D-RAO (L-RAO) magnetizes the magnetite surface along the same direction as the north (south) pole of a magnet \cite{ozturk2022homochirality, ozturk2023magnetization}. Therefore, the coupling between enantiomeric excess and net magnetization is mutually reinforcing, leading to the possibility of a cooperative feedback between chiral molecules and magnetic surfaces, as illustrated in Fig. 4C. 

Thanks to this cooperative feedback, a small natural bias in the net magnetization of magnetic surfaces can be amplified and surfaces with nearly uniform net magnetization can be obtained. These magnetic surfaces can then facilitate spin-selective asymmetric processes with high degrees of selectivity on a persistent basis and ultimately be responsible for breaking the chiral symmetry of biomolecules on early Earth.

\section{Summary}

Ribose-aminooxazoline (RAO) is a central molecule in the prebiotic synthesis of ribonucleotides and its chiral resolution is achieved by its crystallization on magnetized magnetite (Fe\textsubscript{3}O\textsubscript{4}). We obtained enantiopure RAO crystals from a fully racemic solution through a process controlled by electron spin, due to the CISS effect \cite{ozturk2023origin}. Moreover, we demonstrated a mechanism whereby RAO can uniformly magnetize Fe\textsubscript{3}O\textsubscript{4} surfaces, by utilizing the CISS effect in the opposite direction \cite{ozturk2023magnetization}. Thus, our previous work demonstrates that RAO could be a highly plausible chiral compound for achieving homochirality in a prebiotic chemistry network, where natural magnetic minerals, such as magnetite, could act as very effective chiral agents due to the CISS effect (Fig. 4).

In this work, we investigated the propagation of homochirality in a prebiotic network, among its different classes of molecules. We propose that, once homochirality is attained in nucleic acids (i.e. RNA), the chiral information can be efficiently transmitted to other molecular classes - namely, to peptides and then metabolites (Fig. 3). Moreover, we argue that the transmission is unidirectional - propagation of homochirality from amino acids or from metabolites does not seem as efficient and robust. The unidirectional flow of chiral information parallels the flow of genetic information from nucleic acids to peptides in biology, which was established by Crick as the central dogma.

In summary, we have (1) shown a robust mechanism and (2) found a plausible prebiotic compound to attain homochirality, and finally (3) suggested a way of propagating the compound-level homochirality to the network level\textemdash establishing RAO as a promising candidate for unraveling the mystery of biological homochirality's origins.

\begin{acknowledgments}
The authors thank the members of the Simons Collaboration on the Origins of Life and the Harvard Origins of Life Initiative for fruitful discussions that shaped the ideas behind this work. This work was supported by a grant from the Simons Foundation 290360 to D.D.S.\\
\end{acknowledgments}

\section*{Data Availability Statement}

The data supporting the findings of this study were published in two separate works by the authors of this study \cite{ozturk2023origin, ozturk2023magnetization}. The data are avaliable from the publisher and/or corresponding author (S.F.O.) upon reasonable request. No new data were created for this study.

\bibliography{aipsamp.bib}

\end{document}